STOCK MECHANICS: a classical approach

ÇAĞLAR TUNCAY

Department of Physics, Middle East Technical University
06531 Ankara, Turkey
caglart@metu.edu.tr

New theoretical approaches about forecasting stock markets are proposed. A mathematization of the stock market in terms of arithmetical relations is given, where some simple (non-differential, non-fractal) expressions are also suggested as general stock price formuli in closed forms which are able to generate a variety of possible price movements in time. A kind of mechanics is submitted to cover the price movements in terms of classical concepts. Where utilizing stock mechanics to grow the portfolios in real markets is also proven.

## 1. Introduction

Prices in all markets do move in the simplest possible way, either up or down. Yet, predicting the direction of any further step in this one-dimensional movement is a tough subject. Random approach is one of the oldest and widely studied techiques on market predictability.[1] Black-Scholes model stands as a successful method, where it was hypothesised that stock prices perform a geometric Brownian motion, with a constant drift term.[2,3] Recent applications of the random method involves some critiques and refinements.[4-6] As a different approach Mandelbrot used fractals to approximate price graphs deterministically.[7] On the side of modeling of financial markets many other tools and procedures, which are developed to model physical systems, are utilized.[8-16] Due to human causes on the variation of prices, some behavioural approaches are also developed.[17-21] After some observations about the origin of power-law distributions and power-law correlations in financial time series, some more radical mathematical approaches are also worked out.[7, 22-31] On the physics side, one particular interesting model is deduced from magnetism in terms of random magnets in a time dependent magnetic field. This model (as first suggested in [32]) describes the collective behaviour of a set of traders exchanging information, but having all different *a priori* opinions.[33] In a similar context, a relation between magnet spins responding to fluctuations in magnetic field and stock prices responding to fluctuations in demand is put forward and benefited to address the question of how stock prices respond to changes in demand. Many more "similarities and differences between physics and economics"[26] are utilized for predicting the market prices one of which, called the combustion model[34] is found to be very astonishing. The problem there is treated by exploiting an analogy with combustion models occurring in physics and engineering. A simple model described by the concentration of a chemical that obeys the non-linear evolution equation is considered to fit the price data of four stocks from NYSE.

The common aim in all the above and similar works in literature is to make use the predictive power of science, and specifically that of physics, to forecast the markets. Every investor has some initial data about the market and some expectations for the future on which investments will be decided about. So, the present stitutuation taking stage in front of her could be considered as the problem and her decision could be considered as solution.

It is clear that there does not exist experimentations in the present field but just observations. As exactly in case of astronomy where Johannes Kepler did found heavenly laws on the basis of the observations by Tycho Brahe, there might well be existing some simple and general laws within the very domain of financial markets hoped to be discovered soon.

The present paper proposes some new approaches about stock markets specifically relying upon observations. The first section is devoted to a mathematization of the market in terms of simple arithmetic relations. Some simple (non-differential, non-fractal) expressions are also suggested as general stock price formuli in closed forms which are able to generate a variety of possible price movements in time. In the second section a kind of classical mechanics is suggested to cover the price movements in stock markets. This formalism is used to develope some pragmatical applications to be utilized to grow the portfolios realistically. Last section is devoted for further discussions and conclusion.

## 2. Mathematization of Stock Markets

Let's call the set containing all initial data $d_j(t)$ present about a stock market at any time t as D(t).

$$D(t) = \{d_j(t)\} \qquad (1)$$

The corresponding solution set P(t) will contain all decisions to be taken at time t about the stocks in terms of buy or sell or wait;

$$P(t) = \{p_k(t)\} \quad . \qquad (2)$$

So, a function f may be defined between them and the accuracy and relevancy in the solution to the equation

$$f(D) = P \qquad (3)$$

will depend mainly on the relevancy and weights of $d_j \rightarrow p_k$ connections put by the investor, which will determine her potential gains or losses as in PANEL 1.

PANEL 1

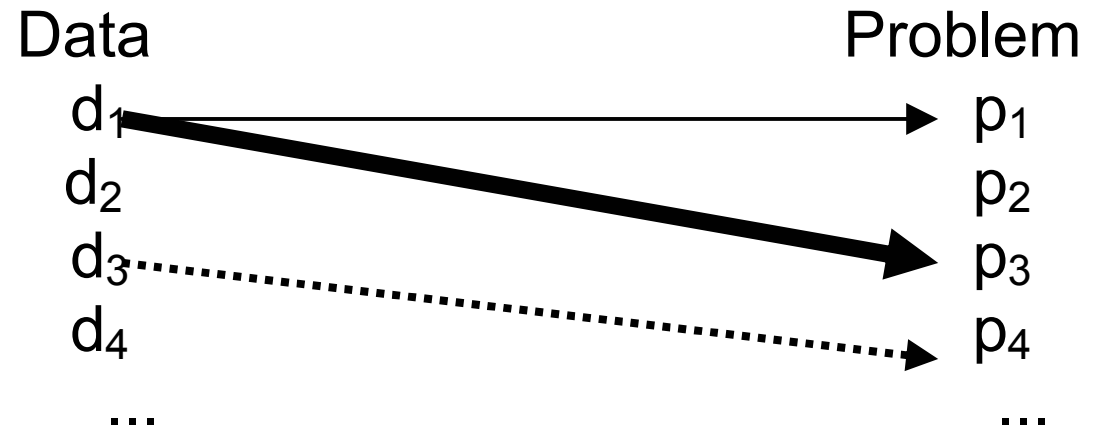

calculation = thinking
$f(d_j) = p_k$

Relevancy and correct weights of $d_j \rightarrow p_k$ connections put by the investors, which determine gains or losses.

Let's suppose that the problem is confined to a single decision $\varepsilon$ to be taken at a moment of time t about how to process on a single stock. In this case $k_{max} = 1$ and $p_1 = \varepsilon$ , where $\varepsilon$ will stand for buy if positive and sell if negative, say. Furthermore, $-1 \leq \varepsilon \leq +1$. Similar inequalities are valid for each investor designated by a super index i and generalization to many i's yields in,

$$0 \leq \varepsilon^i(t) \leq B^i \leq +1 \quad \text{for buy process}$$

and

$$-1 \leq S^i \leq \varepsilon^i(t) \leq 0 \text{ for sell process,} \qquad (4)$$

where $B^i$ and $S^i$ are personal tresholds with different values for different investors, which can be set to +1 and -1 respectevely as defaults in calculations, for simplicity. Definitely, the investor named i will wait in hesitation for $S^i \leq \varepsilon^i(t) \leq B^i$.

If $p^i(t)$ is the portfolio of the investor i, then a market M may be defined as a sum

$$M(t) = \sum_i p^i(t)\, \varepsilon^i(t) \quad , \qquad (5)$$

where the sum is taken over investors (i).

$p^i$ can further be expanded as

$$p^i(t) = \lambda^i(t) + \sigma^i(t) \quad , \qquad (6)$$

and furthermore

$$\sigma^i(t) = \rho^i(t) + \pi^i(t) \quad , \qquad (7)$$

where $\lambda$ and and $\sigma$ are the liquid and the stock amount in portfolio respectively, and $\rho$ is buying cost, $\pi$ is the booked gain on the share before sell.

Therefore, when the period of time $\theta$ previous to the moment of time t is considered, if $\Sigma p^i(t)\varepsilon^i(t) \leq \Sigma p^i(t-\theta)\varepsilon^i(t-\theta)$ then market is said to be sold, and if $\Sigma p^i(t)\varepsilon^i(t) \geq \Sigma p^i(t-\theta)\varepsilon^i(t-\theta)$ then market is said to be bought. In a similar manner, if $\Sigma\lambda^i(t)\varepsilon^i(t) \geq \Sigma\lambda^i(t-\theta)\varepsilon^i(t-\theta)$ then the market is expected to be bought at the moment of t and if $\Sigma\pi^i(t)\varepsilon^i(t) \geq \Sigma\pi^i(t-\theta)\varepsilon^i(t-\theta)$ then the market is expected to be sold at t or around it. Finally all such resulting information will become new data $d_{J+1}$ in the defining Eq. (1) for the next decisive turn, assuming that J was equal to $j_{max}$ for the data set D previous to the moment of time t.

Equation 5 must be generalized for a many-stocks model to cover the reallistic market stituation. For n running over the stock ticks

$$M(t) = \sum_i \sum_n (\lambda^i(t) + \rho^i_n(t) + \pi^i_n(t))\, \varepsilon^i_n(t) \quad . \tag{8}$$

Here it must be pointed out that the liquid term is independent of n. The $\lambda^i(t)$ and $\rho^i(t)$ terms are directly and $\pi^i(t)$ is somewhat indirectly related to tresholds $B^i$ and $S^i$. If $\lambda^i(t) = 0$ (or $\rho^i(t) = 0$) for example, then buy (sell) decision and the corresponding treshold $B^i$ ($S^i$) will become meaningless (unless she exchanges stocks). In the same manner, the negativity of $\pi^i_n(t)$ may prompt her intention for sell to last longer.

Equation 8 can be solved via some simple optimization and iterative methods of linear programming. Outcoming results will extremely be beneficial for the I+1$^{st}$ investor, where $I = i_{max}$ at any time t, provided she exactly knows all the input terms $\lambda^i$, $\rho^i$, and $\pi^i$ at that time for all $i \leq I$. This is practically almost impossible for anybody since she can never reach the official registers of the corresponding individual accounts.

Then one may instead try some hypothetical distribution and density functions for $\lambda^i$, $\rho^i$, and $\pi^i$ which can be borrowed from different branches of science. In order to forecast the stock market realistically, tracing the big portfolio owners such as funds and foundations may suffice, provided the distrubution of $p^i$ is highly inhomegenous over i as averaged in time for sufficiently long term, which is the case in reality. The present scheme can be applied very succesfully with the support of the conventional chartist and fundamentalist analysis techniques and closely follow of legal news.

There are definitely some assumptions made in the above framework. The first of which is that, all investors are close followers, aware of all the relevant information and news. This may be -with some humour- called a "tight-binding approach" borrowing the terms from solid state theory jargon. The second assumption is that, all investors are infinitely reflexive, decisive and do behave without time lags. Which may correspondingly be called as a "pseudo-potential approach" in the same manner. Then, $\varepsilon^i(t)$ will represent tendencies and actions of the investors at the same time. Thirdly, the common aim of all investors is to make profit i.e., to buy the stock shares when they are cheap and sell them when the prices are raised, if the long-term investors making their profits in terms of splits and dividends are excluded. Some "ill-investors" should also be excluded, who may aim at making losses due to mental problems or due to some very special reasons as tax problems, or even to fake the market etc. The third assumption may be called as the "time constant" of the scheme.

Based on the above assumptions, naturally severe inclines and declines in prices may take place consequetively. Since almost all the investors will behave in the same manner, they will either all buy or all sell. As induced by such kind of mechanisms like herding [35-37] repetitions and henceforth some periodicity will be observed in the market.

But the above assumptions can be considered as weak if not entirely false, since they do not reflect the reality perfectly. For example, if all the investors decide to buy (i.e. if $0 \leq \varepsilon^i(t) \leq B^i = +1$ for all i) then who will sell to them ($-1 \leq S^i \leq \varepsilon^i(t) \leq 0$ will be a null inequality, there will be existing no i satisfying it) and vice versa. On the other hand the number of stock shares for each company coded in the market is limited, bourse is not like a "stock pool" or "infinite heat reservoir". Moreover, the investors' profile vary to a wide extend in psychological character as well as their portfolios do in amounts. So, the above assumptions need be relaxed to take into account the randomness in news, investors' individual expectations, their opinions, fears and degrees of gain satisfaction, behaviours etc. To say economically, the variety in $B^i$ and $S^i$ terms in Equation 4 are quite important to induce some portion of randomness on prices and they must be involved in calculational processes. One may note that stock prices are somehow random and do involve some repetitive character.

## A general expression for stock prices and indexes

Combining the three ideas developed in the previous section namely randomness, importance of tresholds and the observed repetitations in buy and sell processes, the present paper proposes a simple expression for the $m^{th}$ day price $\chi_n(t_m)$ of the stock named n, as

$$\chi_n(t_m) = \sum_{q=1}^{m} c_{n,q} \, \mathrm{Sin}(2\pi d_{n,q}) \qquad (9)$$

where, the scalars $c_{n,q}$, and $d_{n,q}$ involve all the randomness and treshold information for the investors sharing the stock named n.

Some early remarks about the present expression may be enumarated as follows:

1) The time t is not selected as a continuous parameter since it designates the $m^{th}$ day of observation. (Stock markets are not continuously open.)
2) The proposed expression involves the history (information about past), in terms of the sum over q.
3) One may obtain another yet similar expression in cosine instead of sine and take any linear combination of the two to extend the generality. In PANEL 2 any arbitrarily chosen graphic out of such almost infinitely many possible ones is displayed, where one branch utilizes sine and the other utilizes cosine. Some of the corresponding scalars are also involved there.
4) The final expression is so general that it may cover all the existing price charts and even the ones that will occur in the future.
5) One may obtain bearish and bullish charts, by playing with the treshold terms.
6) Some more developed charts may also be obtained after some simple manuplations performed in order to involve the intraday fluctuations too as shown in PANEL 3.
7) It is clear that, (due to case 3) here) $\chi_n(t_m)$'s form (at least) a simple algebraic group since multiplication and even division is defined (since the prices do not become zero) between them.

PANEL 2

| Sine (Black) | Cosine (Red) |
|---|---|
| -0,86747 | 0,497482 |
| -1,72382 | 1,013892 |
| -0,87167 | 1,537203 |
| -1,6162 | 0,869607 |
| -1,03935 | 0,052757 |
| -1,79751 | -0,59931 |
| -2,77587 | -0,80621 |
| -1,81975 | -0,51325 |
| -2,81418 | -0,61866 |
| -3,43786 | 0,163019 |
| -3,94993 | -0,69592 |
| -3,89645 | 0,302647 |
| -4,75517 | 0,815105 |
| -5,01573 | 1,780561 |
| -4,11922 | 2,223573 |

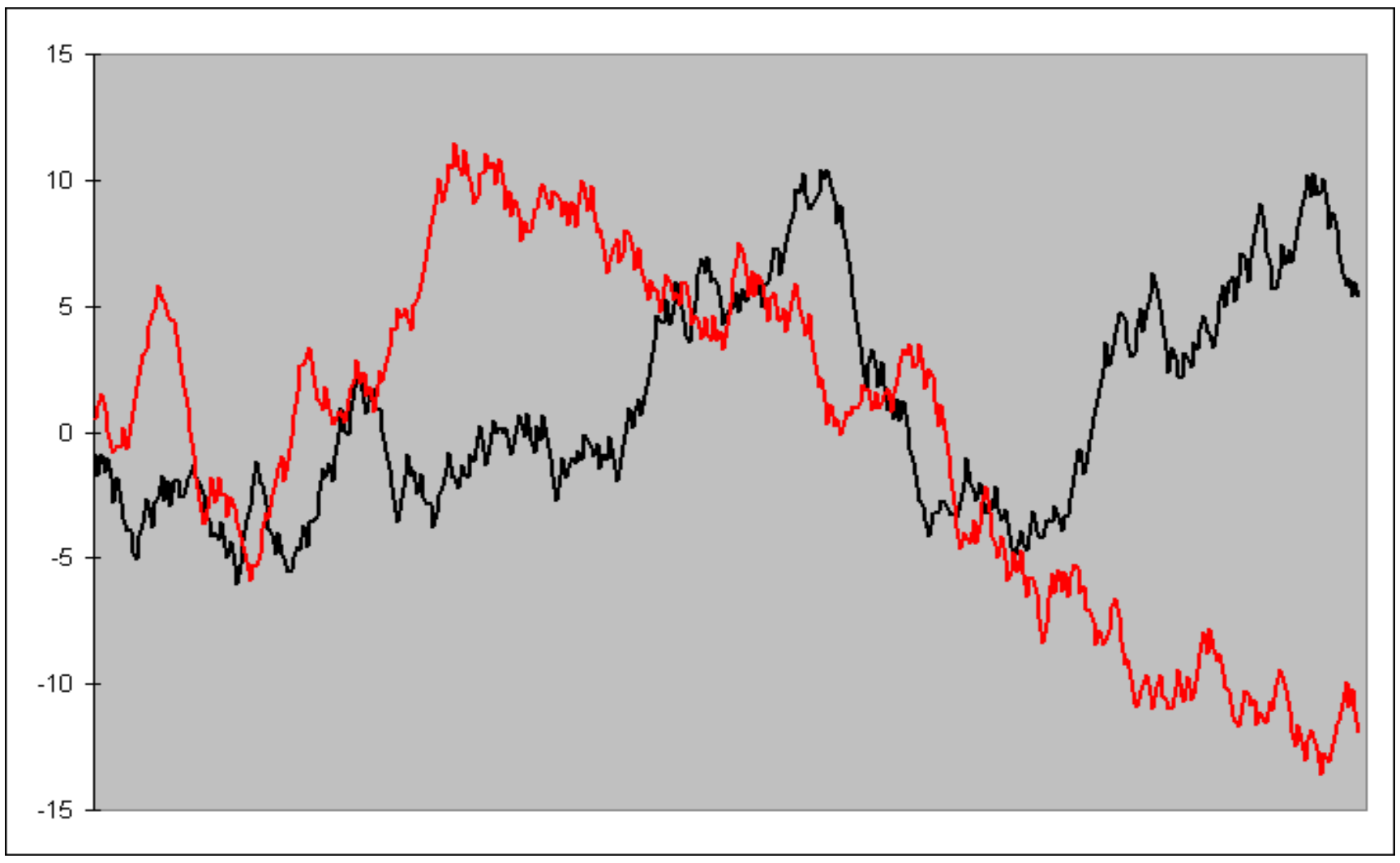

An arbitrarily chosen artificial stock price graphics.

PANEL 3

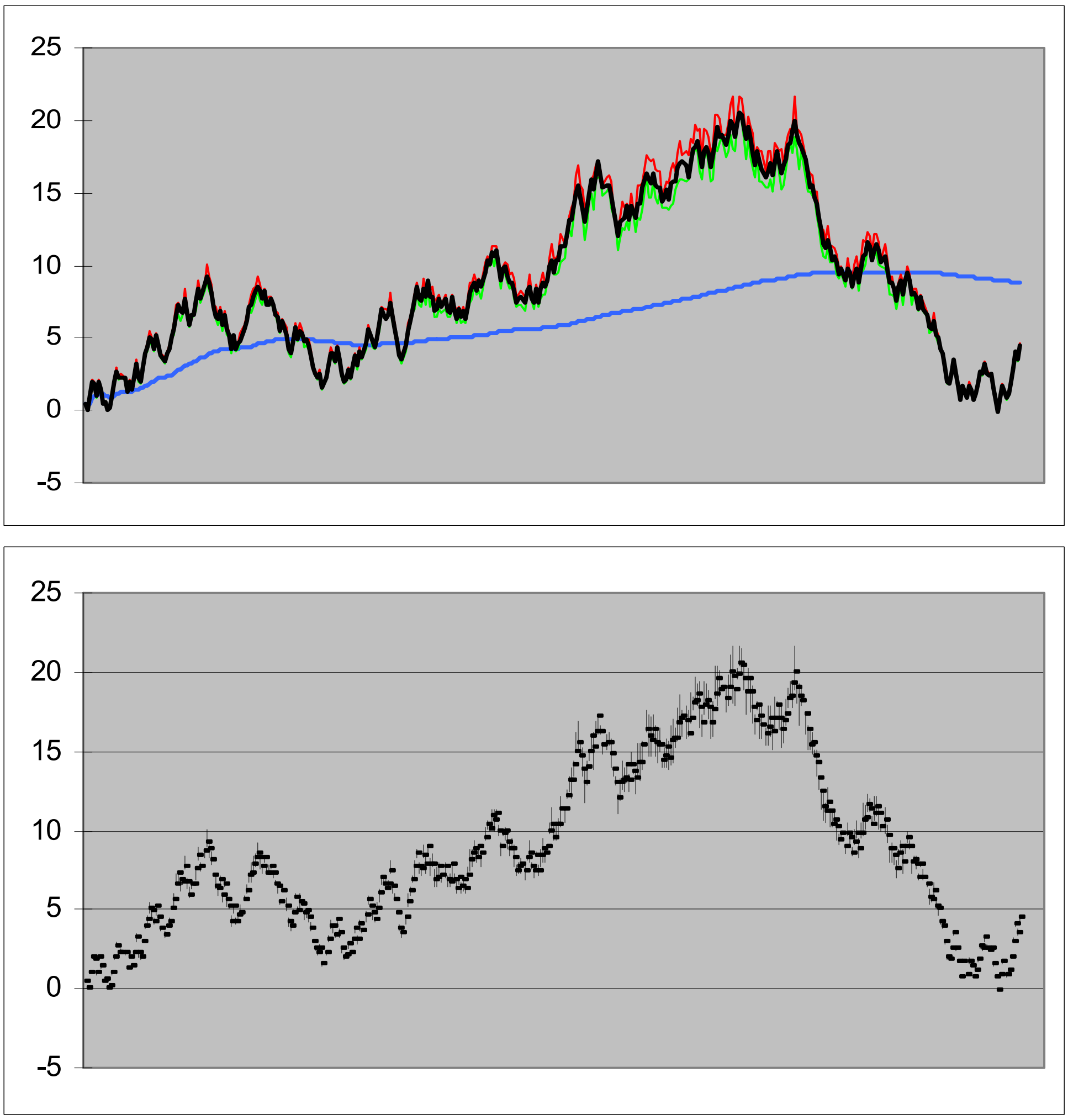

An artificial stock price graphics generated using Equation 9 involving intrinsically the “intraday” fluctuations.

As further remarks on the subject, it can be stated firstly that, any of the infinitely many $\chi_n(t_m)$'s can be produced artificially being independent whatsoever m is, i.e. for infinitely long horizon. Within the same context, any reasonable portion (say, between the days m and m + Δm) of any of $\chi_n(t_m)$ can be considered as an intraday fluctuation of the price too, the extrema of which may also be incorparated into the price charts for any day. Secondly, one may store such artificially produced price graphics (in any number on desire) on a large memory on a computer and search for several aspects involved in them, specifically the relations between the main realms of the graphics and short period and even intraday fluctuations and the treshold terms of Equation 4, the α- and β-correlations etc. to investigate the structure of bearish and bullish markets for example. The results may also be compared with these of data collected from all over the world's real stock markets. Fınally, it is worth to note that, the conventional chartist techniques seem to stay as giving successful results on the present artificial designs. This is quite interesting because chartist techniques were developed statistically over a long (lasting about a century) history where there were real buyers and sellers. Moreover, they are applied each real day by again real buyers and sellers on the aim of making real profits...

## Stock Mechanics

An interdisciplinary and systematic investigation about the science of stock mechanics may be started by observing the existence of the following basic initials:

1) Conservation of the number of shares: The number of stocks belonging to any company coded in any market is fixed. (The splits, dividends, bonuses etc. lay out of the context definitely.)
2) Indivisibility of stocks: Fraction of a share does not exist in any market.
3) Exclusion principle: During any process on any share there can only be one buy and one sell at a time, no more no less.
4) Galilean invariance principle: Shares do not gain or lose value by itself. If there occurs not any important buy or sell process on them, stock price do not change. (Stocks do not move -change price- by themselves. If there is no change of expectation, (falling, rising or horizontal) mood of stocks survives.)
5) Zero acceleration at any time: Balance between buyers and sellers. (See the following subsection.)
6) Applying a force means, doing something to accelerate the prices, i.e., buy or sell heavily. (See the following subsection.)

It is to be underlined that in the above framework, every ınvestor aims at profit during her buy or sell process on any stock and even while waiting.

## Speed, acceleration and force

Observing also that the prices can only make a one-dimensional motion, one may cite a general price value $\chi(t)$ on the positive sense of a one-dimensional axis. Then the time derivative of $\chi(t)$ may be called the speed or velocity, since the direction and senses are well defined. So, $v(t)= d\chi(t)/dt$; which is already called rate of change (ROC) in conventional chartist jargon. The acceleration may also be defined in a very similar way; $a=dv(t)/dt = d^2\chi(t)/dt^2$ .

Moreover, by taking mass equals unity one may define kinetic in a manner similar to physical kinetic energy for the price, i.e., $K= ½ v^2$ where the dimensional unit of K will change from one country to the other; e.g it may be taken as $(\$/sec)^2$ or more properly as $(\$/day)^2$ in USA.

Within the absence of a real particle, no mass (in physical meaning) may exist. Therefore, not any force can be defined presently. Yet some expression to represent the potential may be defined depending upon the stituation taking place in markets.

## Oscillatory motions

In stock markets, extraordinary behaivour in prices are met rarely. Somewhat calm oscillations are more common, especially when there are not big positive or negative expectations. Then, the prices oscillate about an almost horizontal average, which will be designated by $\chi_{av}$. In fact, $\chi_{av}$ may depend on time with a positive slope if a positive expectation is present and with negative slope if a negative one is present. So, the slope of $\chi_{av}$ is determined in parallel to the strength of expectation under consideration. It is natural that, the dips and tips of the corresponding oscillation will behave (in terms of inclination, declination or staying still) accordingly.

When $(\chi_{av} - \chi)$ is large, there occurs a potential to buy since there occurs a potential to make profit out of it. If the depth is high, then the token potential will be high and in any case it will be some function of $(\chi_{av} - \chi)$. So let's call it $U(\chi_{av} - \chi)$. It is an observational fact that, while some investors are buying some others do sell since they hope that the decline in the corresponding price may last for some more days and they may rebuy the same shares cheaper than they sold, or due to fear, or on the aim of exchanging stocks, etc. The reason may differ with respect to the inuniformity within the investor profiles. But some keen investors may notice that there is a buying demand on the token issue. So they may not sell any more, or even they may become new buyers. As a result the number of stocks put on sell decreases in time or increases in price. It is is observed that in such circumstances, the big players of the markets stress the prices by sending some heavy but tricky sell ordinos to little higher prices in order to prompt the other investors for sell. This buying operation in lover prices may last for some days depending upon several other conditions, meanwhile the prices stay almost the same. Then, the speed and accelaration and kinetic become almost zero.

Whenever sufficient buying is performed, an evaluation process takes stage in terms of elavation of prices, then speed and kinetic increase naturally -in the absence of heavy sell or heavy buy ordinos sent- till $(\chi_{av} - \chi)$ becomes negatively large. As $(\chi_{av} - \chi)$ becomes negatively maximum, there occurs a potential to sell and realize the booked profits. In a reverse manner described in the above paragraph, some investors buy these stocks expensively. Furthermore ıt is observed that the big players in such circumstances stress the prices again, but this time by sending some heavy and tricky buy ordinos to little lover prices in order to prompt the other investors for buying. This selling operation may last again for some days during which the prices stay almost horizontal together with the corresponding speed and accelaration and kinetic. Methodologically, not all the stocks at hand are sold at maximum prices, some are witheld to be sold later in order to decline the prices back to suitable and beneficial buy value levels...

This procedure can be summarized mathematically after defining a suitable potential for the stituation alreadily described, as

$$U(\chi_{av} - \chi) = - h (\chi_{av} - \chi)^{\alpha} \quad , \tag{10}$$

where h and α are some parameters to be determined in accordance with reality. It is well known that for α=1 we have a gravitational potential with h equlas to gravitational acceleration on the earth surface, which does not yield any oscillatory motion. Notice that for $\chi_{av} = \chi$ the potential becomes zero, so $\chi_{av}$ may also be defined as the zero-potential-level which is taken conventionally as the earth surface. In order to obtain oscillations, it is clear that α must be equal to 2 and then h becomes as the Hook constant divided by -2.

There is not any evidence for the conservation of the sum of kinetic and potential described above. However, a sinusoidal function may be assumed for oscillatory behaviour in prices.

$$\chi(t_m) - \chi_{av}(t_m) = A\mathrm{Sin}(wt_m + \Phi) = A\cos(wt_m + \Phi') \quad . \tag{11}$$

The expression given in Equation 11 succesfully describes the buy and sell proceses outlined above, with the qualilatively correct $v(t_m)$, $a(t_m)$ etc. In any case, the present scheme may be refined further by considering many and coupled "spring-mass" systems to involve the variety in investor profiles. On the other hand, the angular frequency w can be empirically determined for certain stocks in certain periods of oscillations and hencefort some suitable spring constants may be attained for the corresponding potentials. PANEL 4 displays some oscillatory motions in some stock prices selected from the Istanbul Stock Exchange (ISE) very many similars of which are amply available in every market. The corresponding dynamical parameters, determined empirically as the slope of $\chi_{av}$ (which is equal to $v_{av}$ by definition), amplitude (A in Equation 11) and the Hook constant as the "spring constant" (in units of $(\mathrm{rad/sec})^{-2}$ or more properly in $(\mathrm{rad/day})^{-2}$ since definition for force is not present) are available in PANEL 5. The calculated sinosoidal curves are displayed in PANEL 6 on the aim of any comparisons.

One may also express Equation 11 in terms of the dynamical parameters as

$$\chi(t_m) = \chi_{av}(0) + t_m v_{av} - A\mathrm{Sin}(t_m\sqrt{H}) \quad , \tag{12}$$

where the phase term ($\Phi = \pi/2$) is chosen to have the price at its first oscillatory minimum for the first day ($m=0$, $t_m=0$) of the interval of time which will be considered soon.

# PANEL 4

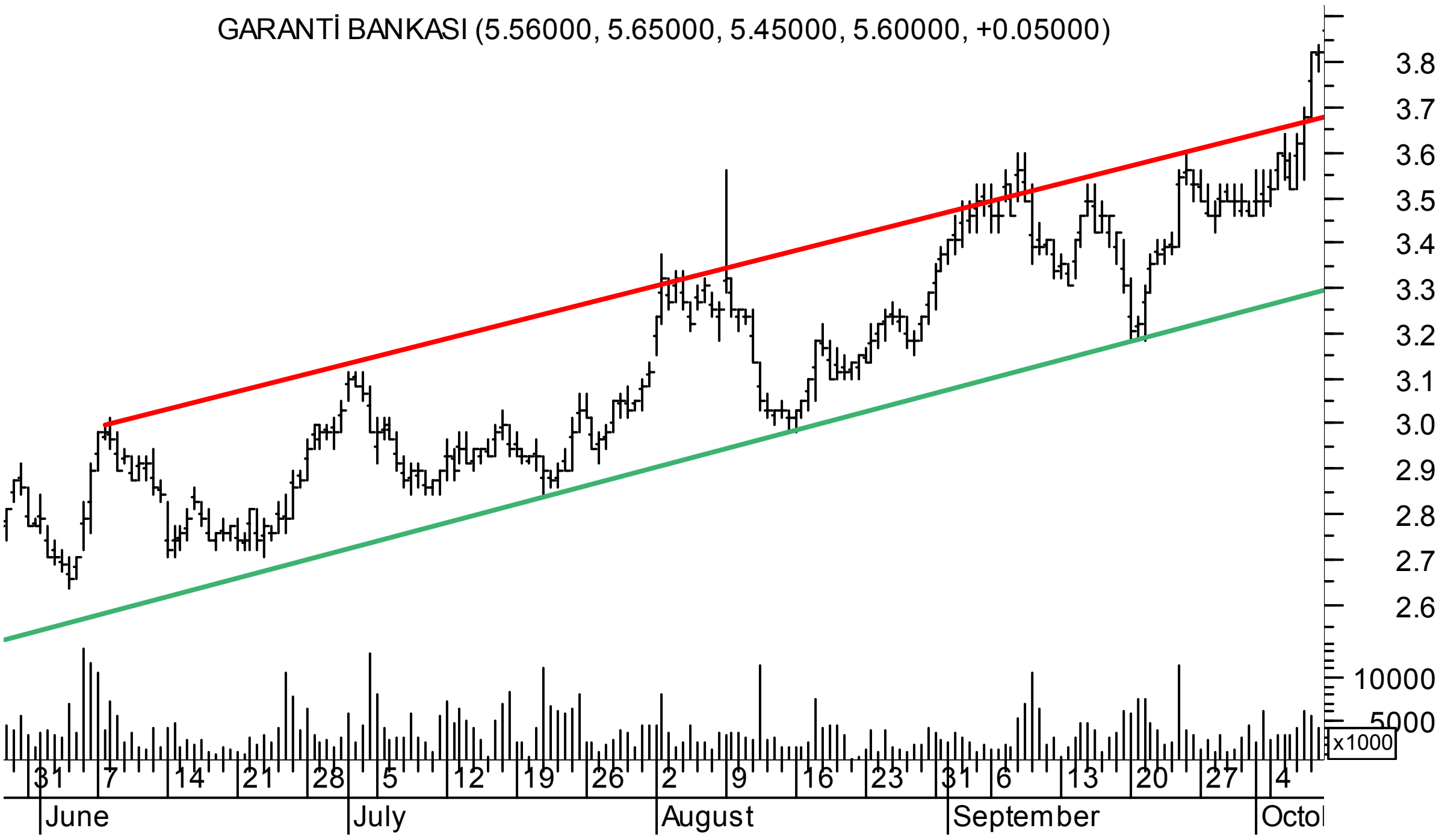
GARANTİ BANKASI (5.56000, 5.65000, 5.45000, 5.60000, +0.05000)
3.8
3.7
3.6
3.5
3.4
3.3
3.2
3.1
3.0
2.9
2.8
2.7
2.6
10000
5000
x1000
31
7
14
21
28
5
12
19
26
2
9
16
23
31
6
13
20
27
4
June
July
August
September
Octol

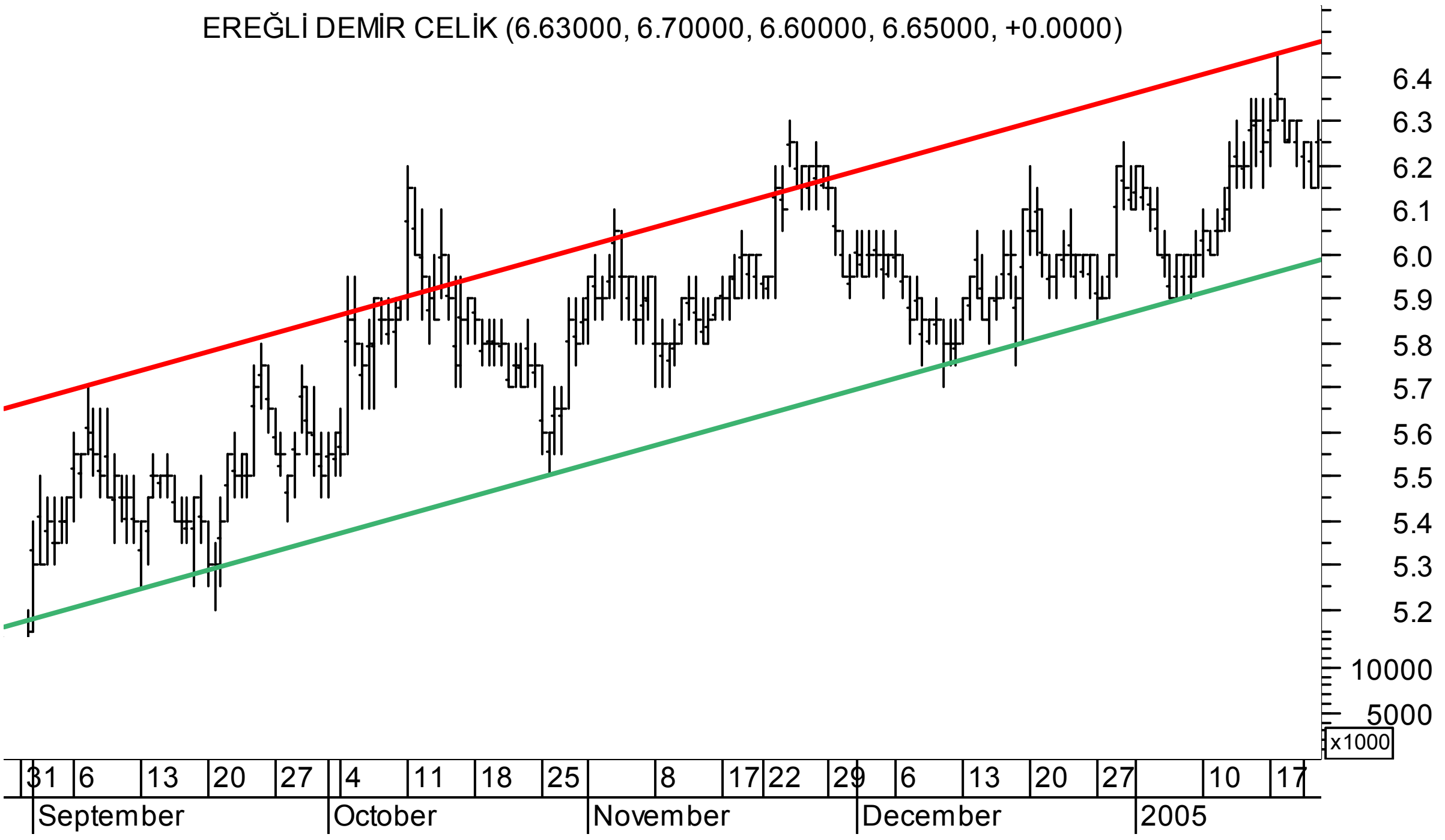
EREĞLİ DEMİR CELİK (6.63000, 6.70000, 6.60000, 6.65000, +0.0000)
6.4
6.3
6.2
6.1
6.0
5.9
5.8
5.7
5.6
5.5
5.4
5.3
5.2
10000
5000
x1000
31
6
13
20
27
4
11
18
25
8
17
22
29
6
13
20
27
10
17
September
October
November
December
2005

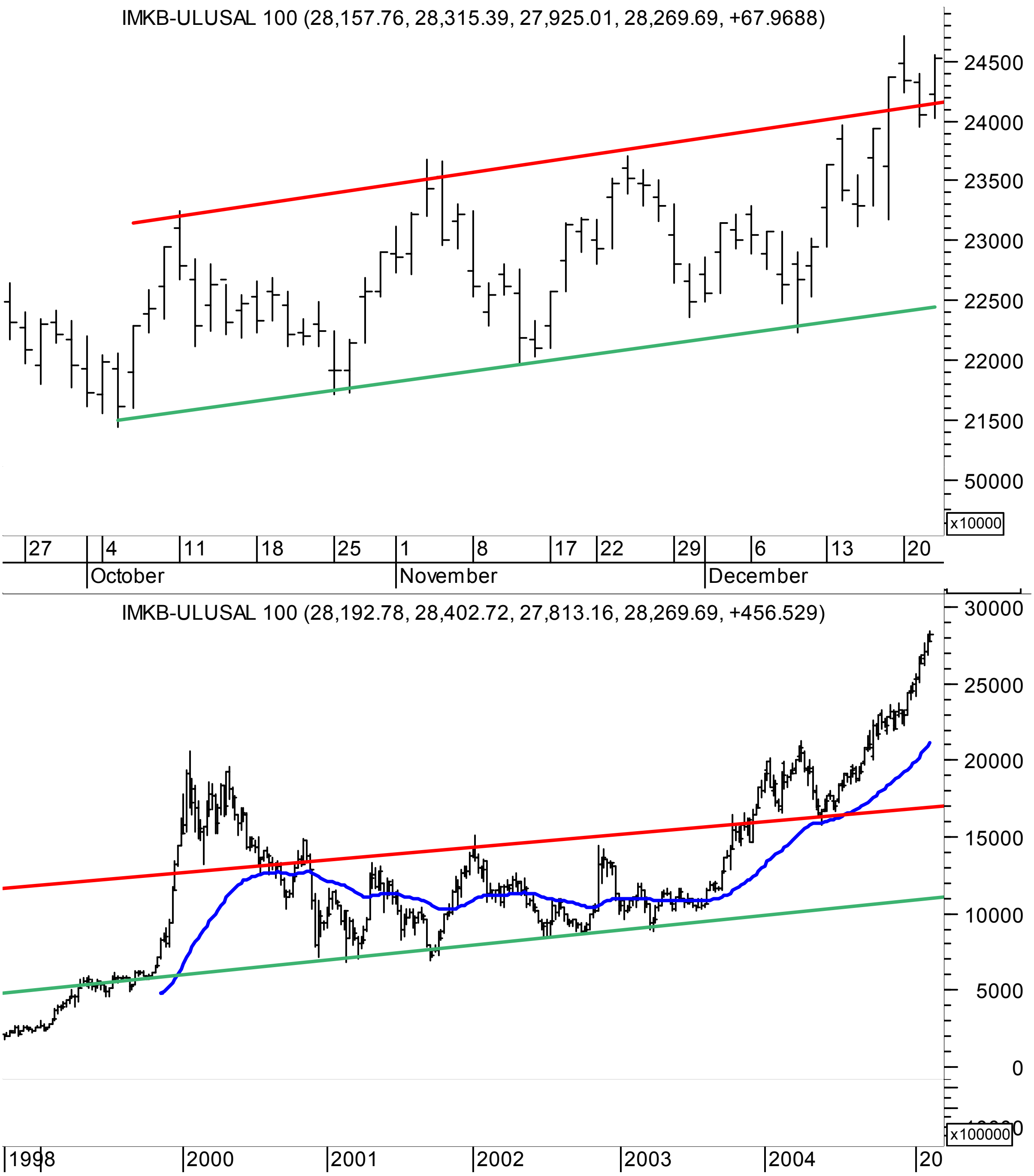

Some oscillatory motions in prices from ISE

Top to bottom: GARAN (a private big bank[38], where the corresponding $\chi_{av}(t_m)$ is neatly upward), EREGL (the biggest iron steel factory in the country, [38]) and the endex ISE-100 (involves the biggest 100 companies in the market). Long range oscillations may even exist as present in the last chart of ISE-100 in weekly unit time basis.

PANEL 5

Dynamical parameters:

| | slope of $\chi_{av} = v_{av}$ | amplitude, A | "spring constant", H |
|---|---|---|---|
| GARAN | 0.0081 YTL/day | 0.044 YTL | 0.081 $(rad/day)^{-2}$ |
| EREGL | 0.0098 YTL/day | 0.250 YTL | 0.044 $(rad/day)^{-2}$ |
| ISE-100 | 0,0182 unit/day | 0.3695 unit | 0.175 $(rad/day)^{-2}$ |

PANEL 6

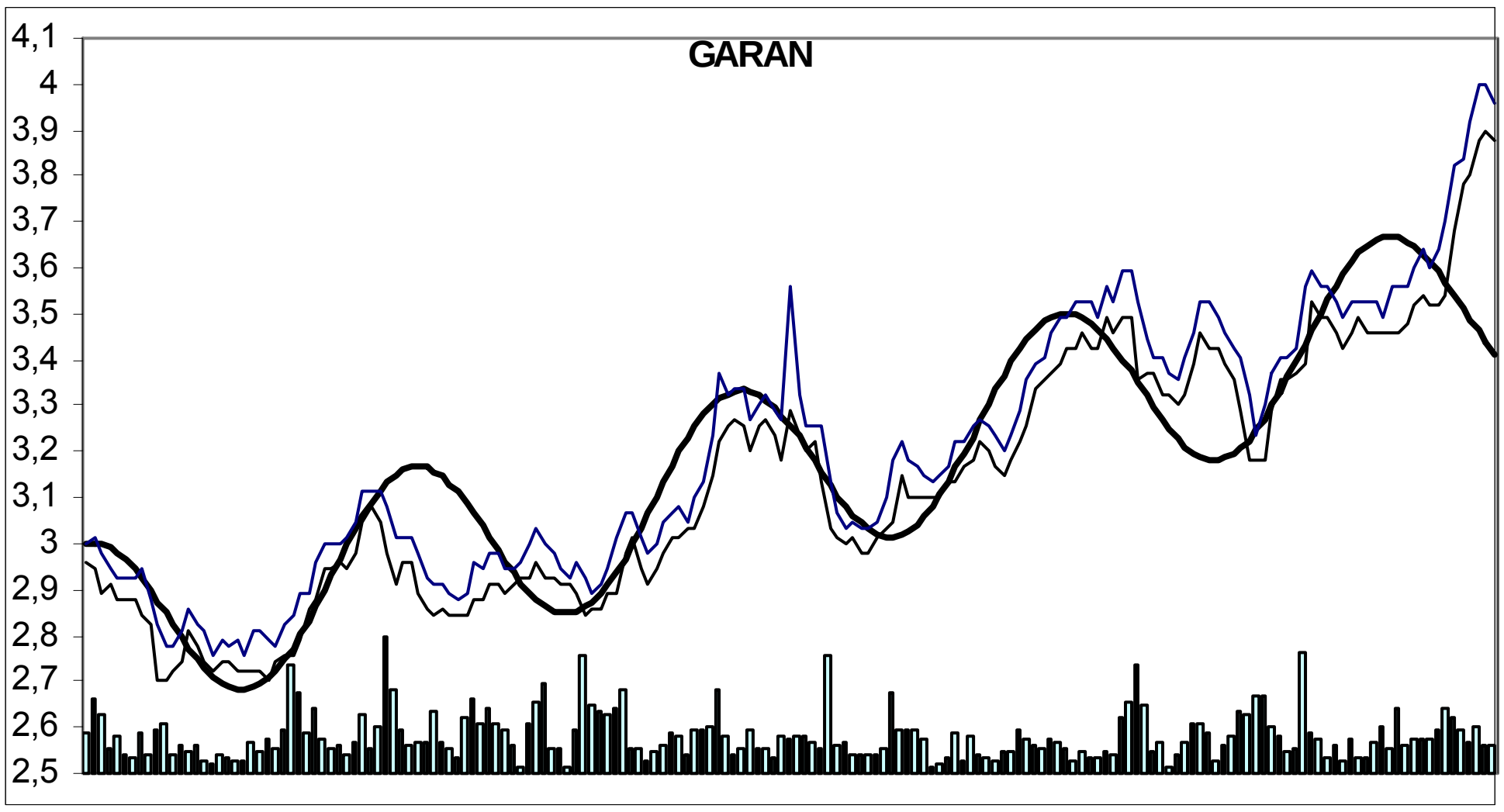

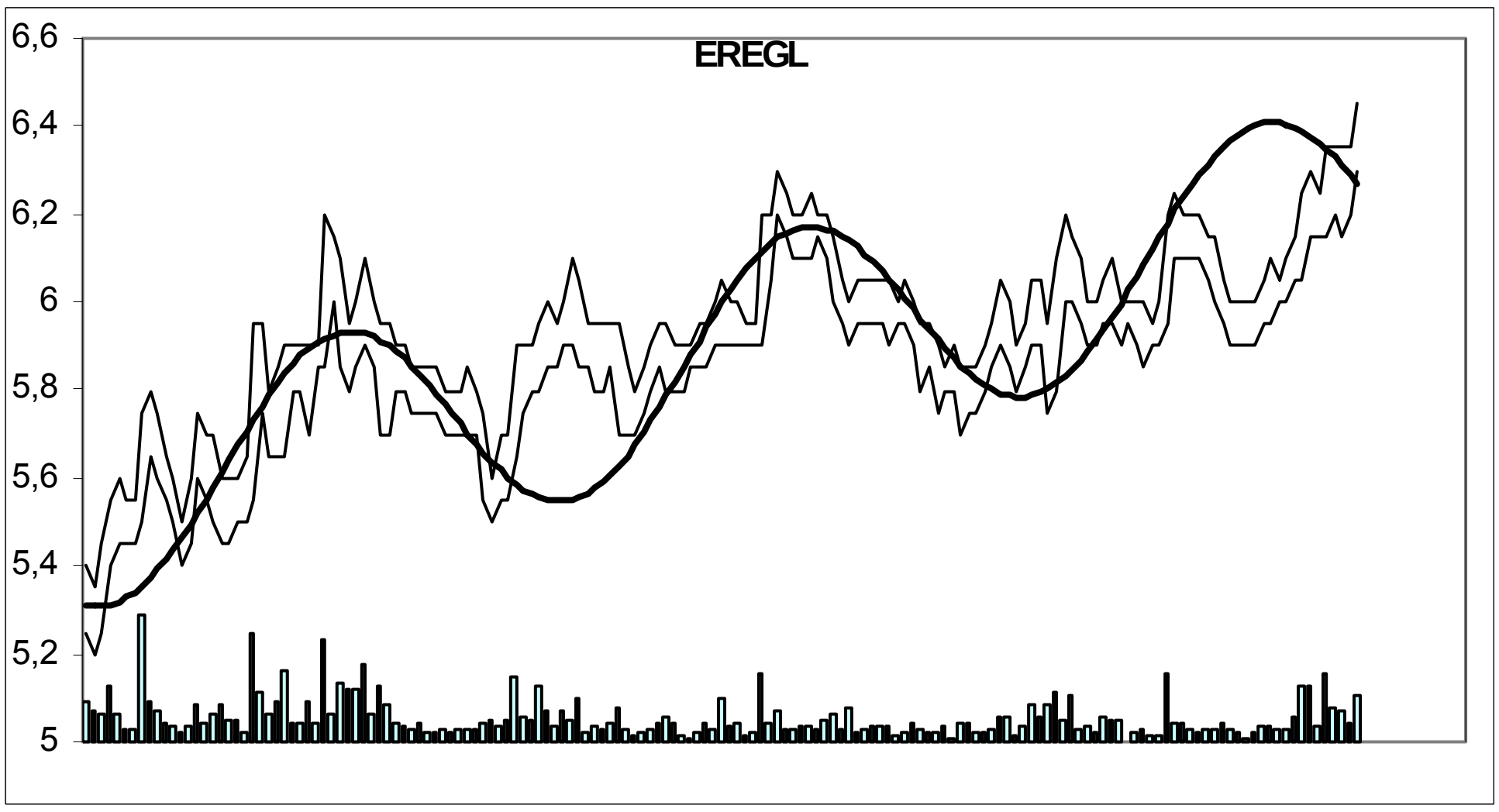

Same curves as in PANEL 4, together with the semi-empirically determined sinusoidals (solid lines), where the upper and lower light lines stand for high and low prices, respectively.

**Oscillatory motions and arithmetical growths in portfolios**

Equation 12 and the dynamical parameters of PANEL 5 are very beneficial for managing grows in portfolios. For short term players stocks with high amplitude and high “spring constant” are advisable, whereas for longer range investors the ones with high $v_{av}$. Note that $v_{av}$ is almost unimportant for the players, because they gain not from the slope of the price but from the oscillations in it. Consider one such player (say, Lady Stomech, as a well experienced person in stock mechanics) observes at the day (m=0, $t_m$=0) the stress-and-buy process (in the morning of 23.June.2004 in the real history) at the stock named GARAN from ISE. Volume is almost doubled (in a half day) with respect to that the whole day before, but the price did not change much. In the afternoon the same day, volume increases by another factor of two and price stays almost still due to heavy sell ordinos lying at slightly higher price levels. (GARAN graphics in PANEL 4) Then, she becomes enthusiastic about buying and sets her B and $\varepsilon$ to +1. After the adjacent a few days of anxiety, she relaxes because of rising the prices with increasing speed. When the speed starts to slow down, about $\pi/2\sqrt{H}=10$ days later she may be alerted for the realization time and soon sets her S and $\varepsilon$ to -1 this time. Then she sells at the beginning of July 2004 with a profit of $\chi(10) - \chi(0) = 10x0.0081 - 0.044Sin(10\sqrt{0.081})$ unit approximately equal to 8.5% of her buying capital. Then the prices start do decline, then declination speeds up, then settles calm again within 10 days. So a full period of oscillations is completed. This time, the price balances at about $\chi(20) = \chi(0) + 20x0.0081 - 0.044$ unit. Her new process of buy at dip and sell at tip will result in the same amount of profit (since the prices are shifted in parallel) but the percentage will be lowered since the buying cost in this new turn is increased, so the ratio becomes smaller. In the third (and may be in the forth, if exists) turn, profit in amount will again be the same and percentange will fall a little more.

What should she do afterwards?... Continue as before? Or try a new opportunity, or stay liquid? It is a tough decision for ordinary people, altough there are just three choises in front. Staying liquid is the riskless choise definitely, but it is out of question here. Trying a new opportunity might be quite beneficial since a new oscillatory term is ready to take place at EREGL at about the middle of September 2004 which is almost the end of the previous oscillatory term in GARAN. So our Lady may prefer to start a new but similar episode there, for new profits.

Since the percentage gains are small here, we may consider Lady Stomech’s (LS) portfolio ($p^{LS}$) growing in each tour at a stock n with a time period of $T_m$ arithmetically, i.e.,

$$p^{LS}(t_m + T_n ) = p^{LS}(t_m) + \Delta p^{LS} \quad , \qquad (13)$$

where $\Delta p^{LS}$ will be a few bigger than $2A_m$ for positive $v_{av}$’s, since the capital $p^{LS}$ does not stay constant but incerases and contributes to gain in amount more and more, provided that total amount is kept in use for buyings in each tour.

Maybe little surprisingly, our cunning lady will not aim at trying any of the above choises but she will continue observing GARAN for a final stroke there. Because, oscillations do not last forever and terminate usually with eihter a crescendo or a decrescendo. Three or four periods each lasting 20 process days on the average add up to about a season which is a long time sufficient in any country to happen something important, politically and economically. Besides, the percentagal gains had already started to cease in potential for big players. So the prices, in general, either break down the lover line passing through the dips, i.e. the minima of the sinusoidals or across up the upper line joining the tips, i.e. the maxima of these. By this means a new era opens in the stocks price history.

The new formation which will take place soon may owe any shape, depending on the new conditions. It may be oscillatory again or a stiff increase or decrease or somehow more eleborate one like shoulder-head-shoulder, or an inverse shoulder-head-shoulder etc...

### Perpendicular motions, rises and falls

Perpendicular motions, i.e. stiff and adrubt rises and falls in prices do not occur oftenly. They are exceptional cases in general met in transitional terms between more steady formations as the oscillations. At the end of the oscillatory period at GARAN discussed in the previous subsection for example, prices acrossed the upper (resistive) line and continued to increase to much higher levels (from where a new oscillation term took place, with a negative slope this time). Lady Stomech, as a consummate person harvested these last and the most ripen and delicious fruits. She were already observing carefully the price and volume relation at the end of the third period when the prices were at dip on the lower (supportive) line, which could very well be broken down soon. Luckily she noticed the volumnous buy process there and immediately recognised that in this tour prices may across the upper limit and go beyond as a transition for the next formation and set her B and $\varepsilon$ to +1, more eagerly this time. She were right, the prices increased by another 2A over the upper line resulting in a final gain at GARAN equal to 4A, where A is the amplitude of the previous sinusoidal oscillation that she had benefited three times. A similar transitional rise can be observed also in weekly ISE chart of the PANEL 4. These kind of transitional increases in prices may be consider as structural perpendicular motions.

Another type of perpendicular increase, which might also be considered as structural formation exists generally just after perpendicular falls. Such adrubt falls occur mostly due to misunderstood news or to events which are corrected soon. When it is recognized that there is no danger or it is removed, the prices ascend towards the previous values. Such fall and rise back formations may be called as elastic falls, some daily and intraday examples of which can be seen in AKBNK chart of PANEL 7. In a similar manner, some elastic rises may occur from time to time, due to mostly positively misunderstood news. In general, adrubt rises are followed by adrubt falls and vice versa. It is worth to note that in such events the potential (Equation 10 with $\alpha$=1) and kinetic are conserved and also their sum resultively.

Falls and rises may also form consecutively, by one trigerring the other as seen in FMIZP chart of PANEL 7 which lasted about two and a half months starting at the middle of October 2004. Notice there that, the minima of falls never sag below (the solid curve which is the moving average of 45 days till it is broken down and then) the horizontal thick line which functions as a supporting level at a value of 370 unit. Secondly, the maxima and so the line joining them i.e., the resistive line decrease regularly. Such a behaviour may be consider as inlastic fall, since some of the initial potential (or kinetic) is lost during every hit with the supporting line. The percentagal loss in potential (or kinetic) determines the slope of the resistive stressing line or vice versa. For inelastic falls $\Delta U/U = tg\xi$, where $\xi$ is the angle between the resistive and horizontal lines measured in the usual counterclockwise manner.

At the end of such a price stressing term, one may expect an abrupt change, the direction of which depends upon the temporary conditions. What is almost certain there is that, the movement continues in the initial direction and does not stop before a distance proportional to the term of causal stressing.

Observing this stock towards the end of December 2004 would definitely be a benefical duty for Lady Stomech for a very nice new year gift for herself. She should buy some FMIZP at a price of 400 and sell them at 600's with a profit of 50% in 20 days.

There remains one more type of perpendicular motion to be considered here, which may be called as completely inelastic fall. One example of this kind of motion is observed in the GOLDAS chart of PANEL 7, where prices falled down and stucked to the level about the horizontal thick line passing over two previous maxima; one at the beginning of September 2004, and one earlier at the mid of November 2002 (the lateral one is not displayed). Notice the volumnous sell at December 2004 which served as distribution of the shares from one hand to the public. Lady Stomech is surely observing this share; unless otherwise the same or at least a considerable amount of it is not accumulated back, she will not buy any lot of it.

**Mechanics of rises and geometrical growths in portfolios**

Continuous stiff falls do occur mostly at deep economic and political crises in the country and stiff rises take place soon after better conditions are recovered. It is one of the rule of thumbs for stock exchanges that extreme growths in portfolios can be achieved at the crisises; "the deeper the crises, the cheaper the prices". Individual companies also may round into crises and recover afterwards or some positively surprising brilliant devolopments may reveal at any time in their history of evolution. All these economic deviations naturally reflect themselves in the corresponding stock prices. In such recovery terms the prices blow up like the bubles speed up extending in volume while ascending in water or rocketships elevating with increasing speed due to fuel burned in their engine motors. (For combustion method, [34].) Correspondingly portfolios may be grown geometrically provided the stocks are choosen correctly and the timing for exchanging stocks are performed in a right way.

Let's consider Equation 10 for perpendicular motions, with $\alpha=1$ naturally. For $h < 0$ one may notice that U decreases with increasing $\chi$. Then one may expect the corresponding kinetic and velocity to increase with increasing $\chi$, which is the case for adrubt rises, as can be seen in the chart of FMIZP for example for the period lasted throughout the whole month of January 2005 (PANEL 7). Assuming the conservation in the sum of kinetics and potential one may deduce $v(t) = ht + v_0$ , where $v_0$ is the initial speed and h plays the role of "negative gravitation". By integrating this equation or from the variations in the sum of kinetic and potential one may obtain the relation between price and its speed (ROC) as

$$v(\chi) = \{C - 2h\chi\}^{1/2} \quad , \qquad (14)$$

where C is some constant taking care of the other constants like $\chi_{av}$, $v_0$, etc. As already noted above that $h < 0$ in Equation 14, and for completely inelastic falls $h > 0$ must be taken.

In both cases of stiff rises and stiff falls considered presently, the undertaking movement stops somewhere for some speed and for some hight in price and the Equation 14 becomes invalid. Afterwards the inverse process takes places, i.e., rise ups turn into fall downs and fall downs turn into rises up due to the change in the conditions between the present (or past) and the forthcoming terms.

In any stiff rise, if Lady Stomech performs a percentagal gain of $\Gamma\ \% = \gamma$, then her portfolio increases as $p^{LS} = p_0^{LS}(1 + \gamma)$. By choosing again a share in the second tour with high $|h|$ (instead of holding the one at hand), she may succeed a new percentagal gain of $\Gamma\ \% = \gamma$. Furthermore she may achieve at finding more new shares in market recovering from crises, which is not impossible since one may expect that at least 10% of the total companies coded in the market may be in such a suitable stituation and about half of them may easily be predicted by following the volumnous accumulations and via fundamental analyses on companies. Then, Lady Stomech's portfolio inflates by

$$p^{LS} = p_0^{LS}\sum_{r=1}^{N}( 1 + \gamma^r ) \quad , \qquad (16)$$

where the sum is over tours (r), N is the total number of tours taken by exchanging the shares. It is worth to note that one may manage to grow her portfolio in multiples of 2.9, 4.8, 8.2 for $\gamma$=0.3 and in multiples of 5.1, 11.4, 25.6 for $\gamma$=0.5 in four, six, eight tours, all respectively. Especially in recovery eras while the prices were ascending from their long term minima, it is quite easy to find many shares promising 30% and %50 gains quickly. Lady Stomech had processed even in later dates with %50 at FMIZP in January 2005, as considered previously.

Definitely, one may utilize |h| grafically for choosing the best of the shares available at each buying turn too, since 1/|h| equals to the curvature of the prices just before starting to rise; the smaller the curvature is the speeder and higher the rise is. Some curvatures determined semi-emprically are -0.0127, 0,0009, 0,0190 for AKBNK, FMIZP, and GOLDAS respectively all in $DAY^2$/(currecy unit), where negative value stands for the fall. In PANEL 8 the corresponding curves obtained utilizing Eq. (14) and related analysis in a semi-empirical way are displayed on the realistic price charts, where the close fit may be noticed. Furthermore, the same semi-empirical curves may be used to decide about the right time for selling i.e., setting $\varepsilon$ to -1. It is the time when the difference between them is much bigger than the average value.

PANEL 7

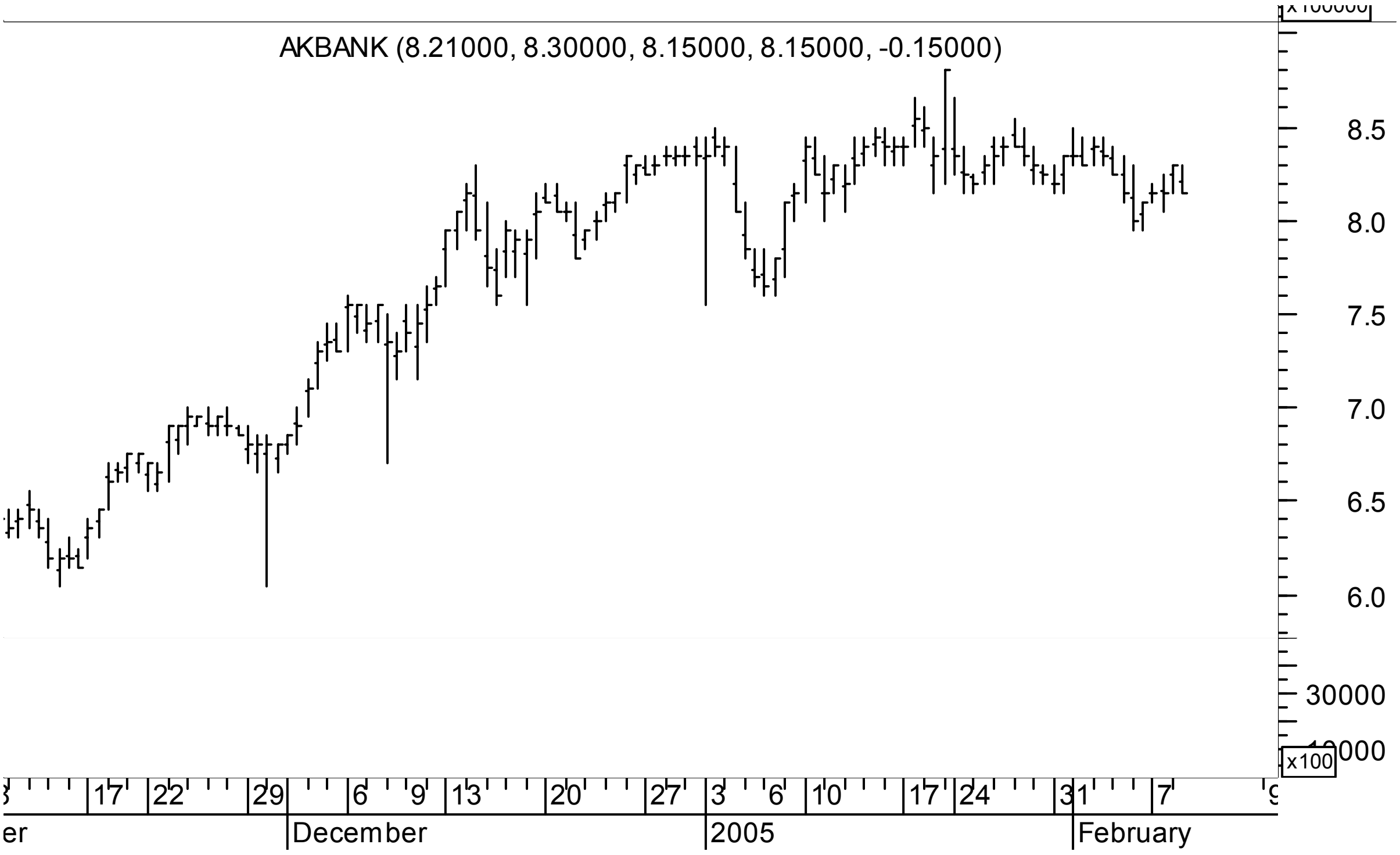

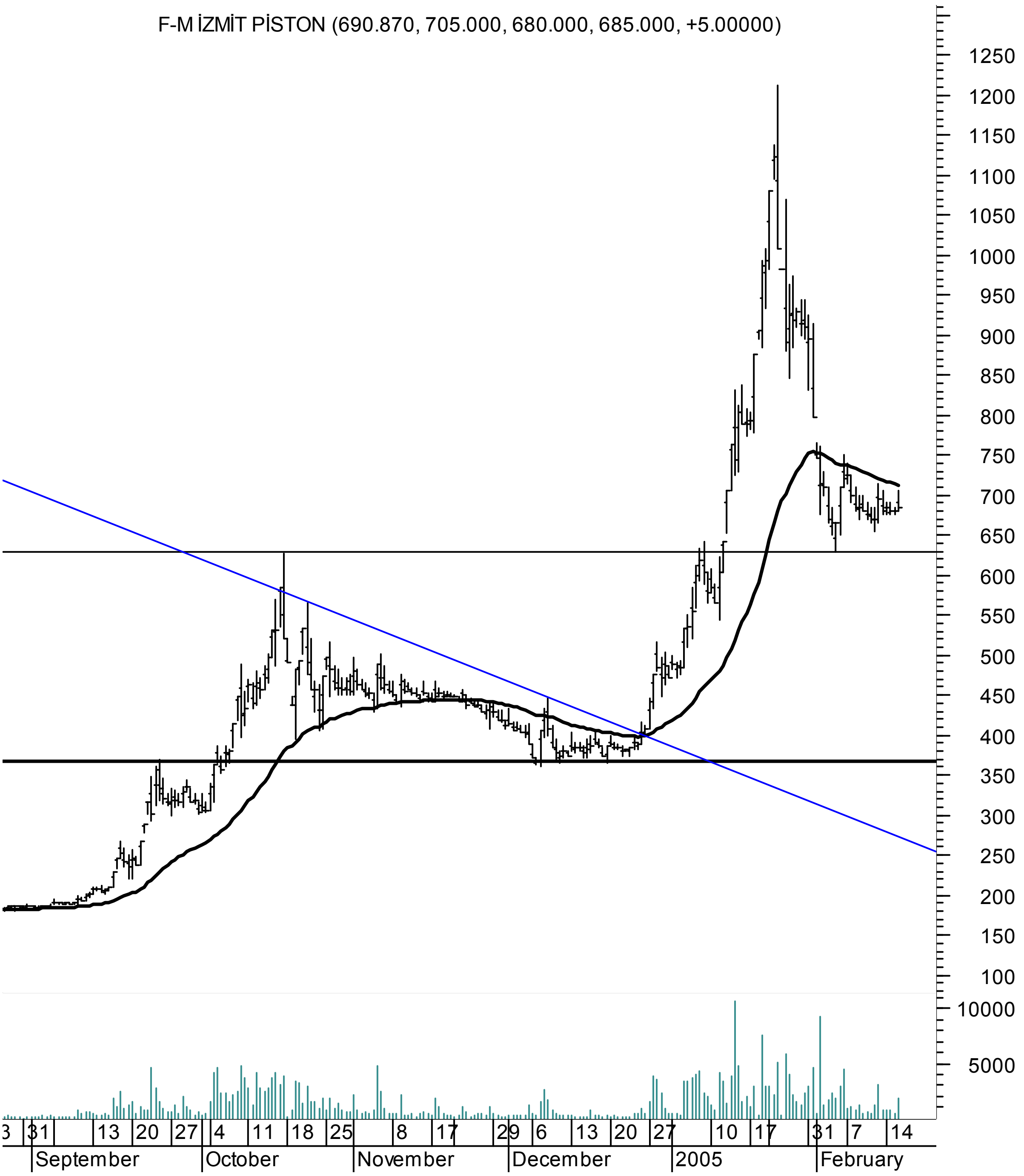
F-M İZMİT PİSTON (690.870, 705.000, 680.000, 685.000, +5.00000)
1250
1200
1150
1100
1050
1000
950
900
850
800
750
700
650
600
550
500
450
400
350
300
250
200
150
100
10000
5000
3
31
13
20
27
4
11
18
25
8
17
29
6
13
20
27
10
17
31
7
14
September
October
November
December
2005
February

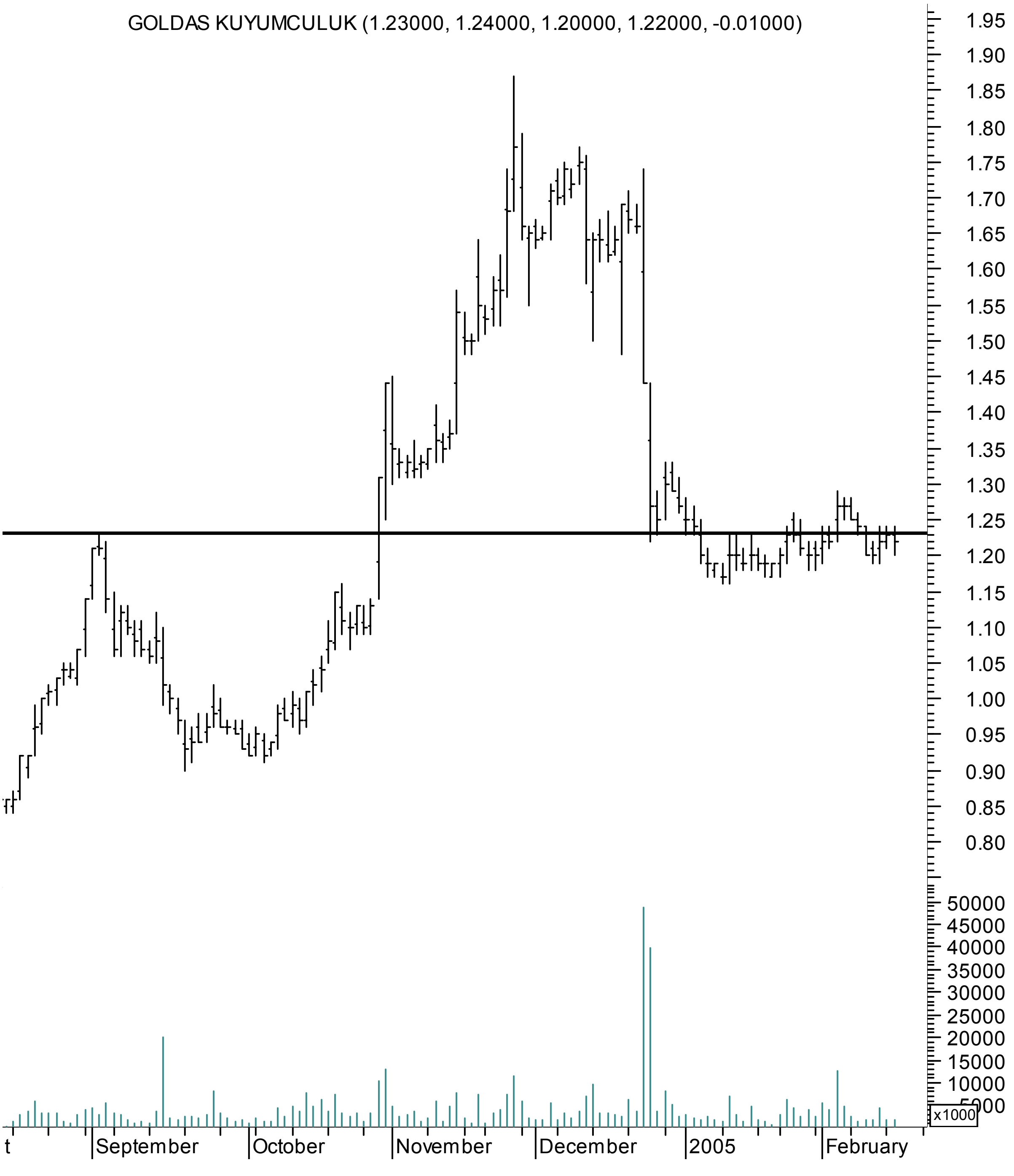

Typical perpendicular motions

From top to bottom: AKBNK (a big private bank[38]), FMIZP (an industrial firm[38]), and GOLDAS[38].

PANEL 8

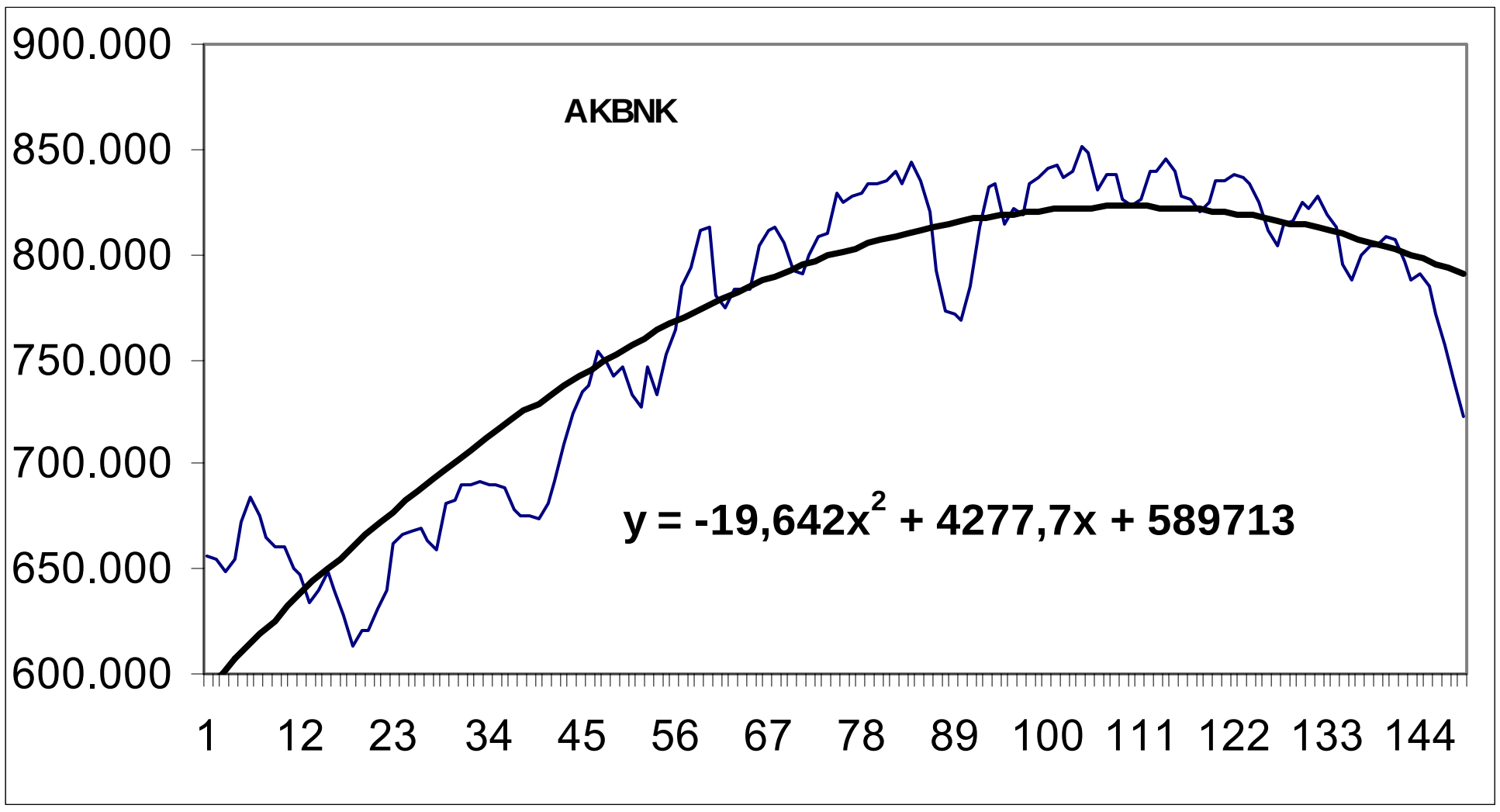

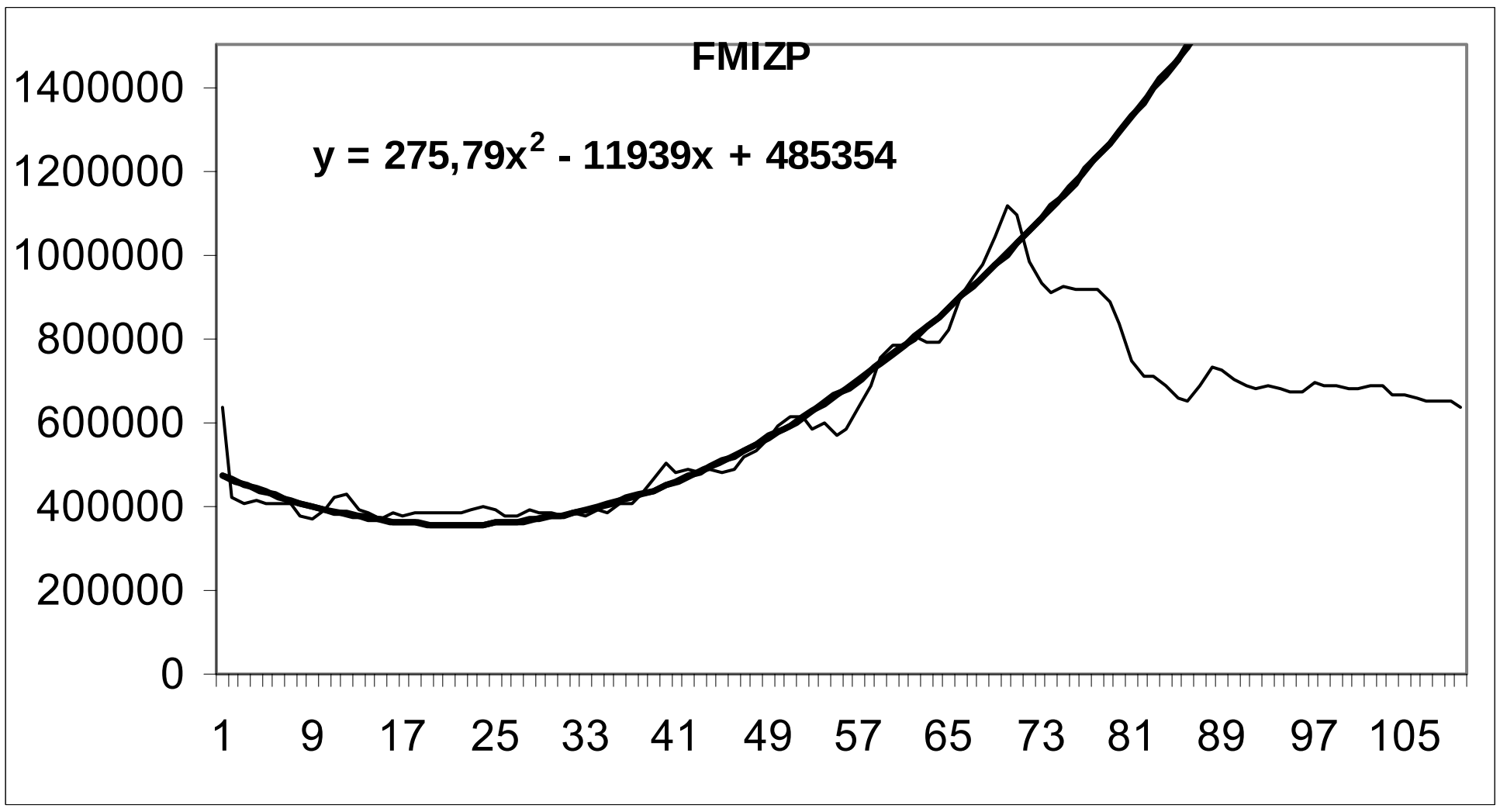

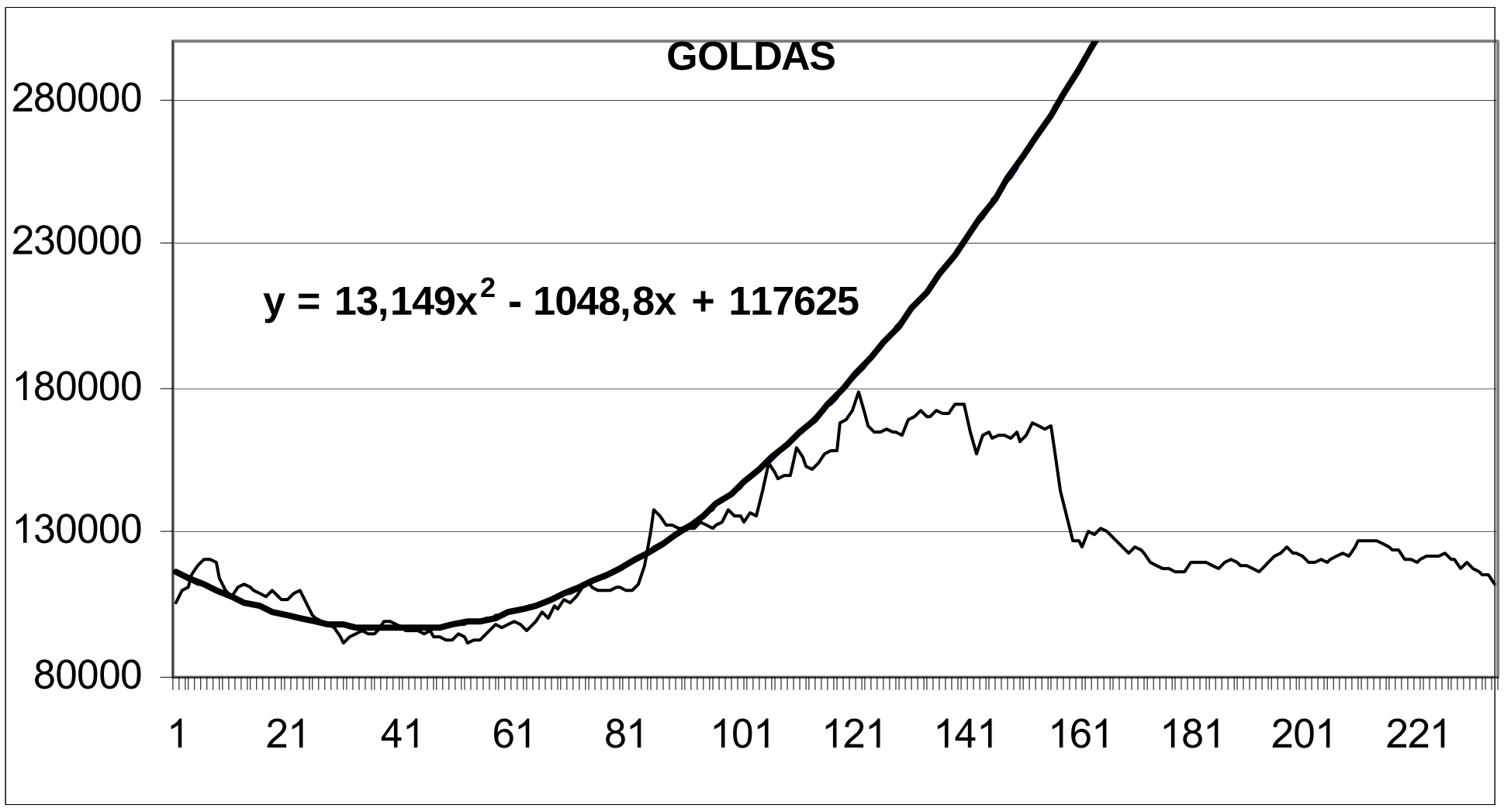

AKBNK, FMIZP, and GOLDAS (as in PANEL 7) and the semi-empricaly determined fitting curves (solid lines) which may be benefited for buy and sell processes.

## 3. Discussions and Conclusion

The present formalism was focused mainly on the maximization of profit in terms of efficient processes, where the principle of minimization in risk as a counterpart of profit was considered implicitly. Stock mechanics is able to determine a variety of ways to choose the profitable shares and timing in buy and sell processes succesfully. However due to the fact that future can not be foreseen in absolute manner, “minimization of risk” might not be considered as a well defined subject. How, for example, Lady Stomech would be sure that any catastrophe will not take place while holding shares and carry them for some longer time to sell at tips determined via stock mechanics? There does not exist any definite answer to this question. However, all types of catastrophes may exist at any time anywhere, they are parts of human lifes. Nobody commits suicide due to the possibility of their occurances. Similarly, the risk of general crashes is taken initially for trading in markets, and it can never be equated to zero. The meaning of “minimization of risk” can be defined beyond this limit.

Stock markets, as any other financial markets are neither completely indeterministic nor deterministic. They are partially predictable. They involve some “rules of thumb”, but there can not be any “golden rule” valid forever. First of all, determinism is a consequence of causality, which is not valid for financial markets, since (even if the potential affects were well determinable) the causes of the events are somehow random. What can be predicted more succesfully are the normal eras between drastic and effectual occurances. Secondly, the effects of occurances on the markets (i.e. on the investors) may be estimated to some extent, provided that the cause-result connections and the weights there are put correctly. Correctness and precision are almost vital in such cases when the risk distribution in the portfolio is considered.

In the present respect, trying to forecast the crashes is one side of the problem. The other side maybe taken as being well prepared for them. The portfolio distribution over the liquid and stock items (incuding the inner distribution over the shares) are to be decided about in accordance with the foreseen probabaility of risks. Where, cause-result connections and the corresponding weights (Eqs. (1-3), and PANEL 1) are worth to be put correctly on the aim of decreasing the potential damages and thus, increasing the overall profits. Note that, the profits can be increased under a certain risk.

Equation 8, and the analysis in the paragraph preceding it, may be benefited on several purposes. Lady Stomech may pick, for example the accumulated shares (in terms of comparably high $\rho$ values) of the market and choose out of them the most suitable ones for herself. Such “bought” shares are known to be resisting against general recessions and also to be recovering quickly if they do relax for any reason. She is able to make a clear distinction between elasticly falled shares and inelasticly falled ones (even in abnormal terms), and sell the laterals if withelding some to buy the primal ones. This is another field where the stock mechanics may be utilized, now, to cure the damages of crashes.

In normal eras, Equation 8 and the related analysis may also serve as a guide for trading purposes too. If for a specific share (n), $\pi^i_n(t_m)$ ($= \chi_n(t_m) - \rho^i_n(t_m)$) has a negatively large value as averaged over all of its (at least big) carriers (i), then it means that this share is potentially beneficial to buy at the day of $t_m$. Similarly if the same $\pi$ term is positively large at another time, for example $\Delta m$ days later, it means that the same share is potentially beneficial to sell then. Such stituations are regularly met at oscillatory dips and tips respectively.

In the present formalism oscillations in prices were worked out in terms of an hypothetical spring-mass system. One may easily substitute it by the electrical resistance-capacitance analog, as another model. Some more elobarate formations as the elastic and inelastic falls and rises are also studied, where transitional formations are treated to some extent. There exist some more formations in price charts, such as double dips, double tips, shoulder-head-shoulders and iverses of them, etc. These are all worth to analyse from the stock mechanical viewpoint. They may very well involve deep structures as some forms of wave packets. Creations, annihilations and transmutations of formations are very crucial subjects to be studied in any attempt about predicting exchanges. At this point, all the real data gathered from all the real markets and the artificial ones to mimic them (as obtained via Equation 9, and via other methods, some of which are refferenced in the opening section of the present paper), and even simulative ones may be benefited to explore any deeper structure(s) in terms of conservations and the -probable- law(s) governing their transmutations, creations, and annihilations.

**Acknowledgement**
The author would like to thank to Prof. Şakir Erkoç for his friendly help and careful reading of the manuscript.